\documentclass[aps,prx,showpacs,floatfix,onecolumn,superscriptaddress,longbibliography]{revtex4-2}
\usepackage[normalem]{ulem}
\usepackage{graphicx} % Required for inserting images
\usepackage{float}
\usepackage{color}
\usepackage{bm}
\usepackage{hyperref}
\usepackage{todonotes}
\usepackage{verbatim}
\usepackage{soul}
\usepackage{glossaries}
\usepackage{sidecap}
\usepackage{hyperref}% add hypertext capabilities
\hypersetup{
    colorlinks=true,
    linkcolor=blue,
    filecolor=magenta,
    citecolor=blue,
}
\usepackage{orcidlink}
\usepackage{titlesec}
\def\Q{\ensuremath{\mathbf{Q}}}
\def\Qp{\ensuremath{\mathbf{Q}_{\parallel}}}
\def\k{\ensuremath{\mathbf{k}}}
\def\TN{\ensuremath{T_{\mathrm{N}}}}
\def\LOMS{La$_2$O$_3$Mn$_2$Se$_2$}

\newacronym{FM}{FM}{ferromagnetic}
\newacronym{AFM}{AFM}{antiferromagnetic}
\newacronym{REXS}{REXS}{resonant elastic x-ray scattering}
\newacronym{RIXS}{RIXS}{resonant inelastic x-ray scattering}
\newacronym{XAS}{XAS}{x-ray absorption spectrum}
\newacronym{FWHM}{FWHM}{full-width at half-maximum}
\newacronym{2D}{2D}{two-dimensional}
\newacronym{3D}{3D}{three-dimensional}
\newacronym{TM}{TM}{transition-metal}
\newacronym{ED}{ED}{exact diagonalization}
\newacronym{ARPES}{ARPES}{angle-resolved photoemission spectroscopy}
\newacronym{CL}{CL}{circular-left}
\newacronym{CR}{CR}{circular-right}
\newacronym{LH}{LH}{linear-horizontal}
\newacronym{LV}{LV}{linear-vertical}
\newacronym{ATS}{ATS}{anisotropic tensor susceptibility}
\newacronym{LSWT}{LSWT}{linear spin wave theory}
\newacronym{DFT}{DFT}{density functional theory}
\newacronym{DM}{DM}{Dzyaloshinskii-Moriya}
\newacronym{CNN}{CNN}{convolutional neural network}

\begin{document}

\title{Resonant inelastic X-ray scattering reveals d-wave altermagnetism}

\title{$d$-wave altermagnetism revealed by resonant inelastic X-ray scattering}

\author{Guangkai Zhang}
\affiliation{Beijing National Laboratory for Condensed Matter Physics, Institute of Physics, Chinese Academy of Sciences, Beijing 100190, China}
\affiliation{Department of Physics, Shanghai Normal University, Shanghai 200234, China}

\author{Yuehong Li}
\affiliation{Department of Physics, The Chinese University of Hong Kong, Shatin, Hong Kong, China}
\affiliation{State Key Laboratory of Quantum Information Technologies and Materials, The Chinese University of Hong Kong, Shatin, Hong Kong, China}

\author{Xubin Ye\orcidlink{0000-0002-5739-8318}}
\affiliation{Beijing National Laboratory for Condensed Matter Physics, Institute of Physics, Chinese Academy of Sciences, Beijing 100190, China}

\author{Vincent C. Morano\orcidlink{0009-0002-1649-0887}}
\affiliation{PSI Center for Neutron and Muon Sciences, Paul Scherrer Institute, 5232 Villigen PSI, Switzerland}

\author{Sze Tung Li\orcidlink{0009-0000-3280-6752}}
\affiliation{Department of Physics, The Chinese University of Hong Kong, Shatin, Hong Kong, China}
\affiliation{State Key Laboratory of Quantum Information Technologies and Materials, The Chinese University of Hong Kong, Shatin, Hong Kong, China}

\author{Jaewon Choi\orcidlink{0000-0003-4616-4345}}
\affiliation{Diamond Light Source, Harwell Campus, Didcot, Oxfordshire, UK}

\author{Rebecca Scatena\orcidlink{0000-0002-3500-1455}}
\affiliation{Diamond Light Source, Harwell Campus, Didcot, Oxfordshire, UK}

\author{Shuai Tang}
\affiliation{Beijing National Laboratory for Condensed Matter Physics, Institute of Physics, Chinese Academy of Sciences, Beijing 100190, China}
\affiliation{School of Physical Sciences, University of Chinese Academy of Sciences, Beijing 100049, China}

\author{Maocai Pi}
\affiliation{Beijing National Laboratory for Condensed Matter Physics, Institute of Physics, Chinese Academy of Sciences, Beijing 100190, China}
\affiliation{School of Physical Sciences, University of Chinese Academy of Sciences, Beijing 100049, China}

\author{Mengqi Ye}
\affiliation{Beijing National Laboratory for Condensed Matter Physics, Institute of Physics, Chinese Academy of Sciences, Beijing 100190, China}

\author{Mirian Garcia-Fernandez\orcidlink{0000-0002-6982-9066}}
\affiliation{Diamond Light Source, Harwell Campus, Didcot, Oxfordshire, UK}

\author{Alessandro Bombardi\orcidlink{0000-0001-7383-1436}}
\affiliation{Diamond Light Source, Harwell Campus, Didcot, Oxfordshire, UK}

\author{Xiaomei Qin}
\affiliation{Department of Physics, Shanghai Normal University, Shanghai 200234, China}

\author{Zhao Pan\orcidlink{0000-0002-8693-2508}}
\affiliation{Beijing National Laboratory for Condensed Matter Physics, Institute of Physics, Chinese Academy of Sciences, Beijing 100190, China}

\author{Daniel G. Mazzone\orcidlink{0000-0002-0421-0625}}
\affiliation{PSI Center for Neutron and Muon Sciences, Paul Scherrer Institute, 5232 Villigen PSI, Switzerland}

\author{Qisi Wang\orcidlink{0000-0002-8741-7559}}
\email{qwang@cuhk.edu.hk}
\affiliation{Department of Physics, The Chinese University of Hong Kong, Shatin, Hong Kong, China}
\affiliation{State Key Laboratory of  Quantum Information Technologies and Materials, The Chinese University of Hong Kong, Shatin, Hong Kong, China}

\author{Yi Lu\orcidlink{0000-0001-9445-641X}}
\email{yilu@nju.edu.cn}
\affiliation{National Laboratory of Solid State Microstructures and
Department of Physics, Nanjing University, Nanjing, China}
\affiliation{Collaborative Innovation Center of Advanced Microstructures, Nanjing, China}

\author{Yao Shen\orcidlink{0000-0003-4697-4719}}
\email{yshen@iphy.ac.cn}
\affiliation{Beijing National Laboratory for Condensed Matter Physics, Institute of Physics, Chinese Academy of Sciences, Beijing 100190, China}
\affiliation{School of Physical Sciences, University of Chinese Academy of Sciences, Beijing 100049, China}

\author{Youwen Long\orcidlink{0000-0002-8587-7818}}
% \email{ywlong@iphy.ac.cn}
\affiliation{Beijing National Laboratory for Condensed Matter Physics, Institute of Physics, Chinese Academy of Sciences, Beijing 100190, China}
\affiliation{School of Physical Sciences, University of Chinese Academy of Sciences, Beijing 100049, China}

\date{\today}

\maketitle

\textbf{Altermagnetism defines a third fundamental class of collinear magnetic order, featuring compensated magnetic moments with antiparallel spin alignment, yet lifted Kramers degeneracy without the need for relativistic spin-orbit coupling ~\cite{Smejkal2022Conventional,Smejkal2022Emerging,Jungwirth2026Symmetry,Cheong2025Altermagnetism}. Its ability to host spin-polarized electronic bands and unconventional chiral magnons makes it a promising platform for functional materials~\cite{Bai2024Altermagnetism,Song2025Altermagnets,Zhang2025Theory}. However, experimental verification has proven challenging; while circular dichroism in resonant inelastic X-ray scattering (RIXS) has been suggested as a signature of chiral magnons~\cite{Takegami2025Circular,Jost2025Chiral,Biniskos2025Systematic,Channagowdra2025Bialtermagnetism}, it remains controversial whether this effect is an intrinsic property of altermagnetism or an artifact of experimental geometry~\cite{Biniskos2025Systematic,Furo2025Theory}. In this work, we resolve this debate and provide unambiguous experimental evidence of $d$-wave altermagnetism in the strongly correlated Lieb-lattice magnet \LOMS{}. The RIXS spectra exhibit a $d$-wave-symmetry circular dichroism in the magnetic excitations that vanishes in the paramagnetic phase. Through RIXS-operator symmetry analysis and exact-diagonalization calculations, we prove that the observed dichroism is a direct consequence of altermagnetic symmetry constraints, independent of magnon branch splitting. Our results provide definitive evidence for the experimental realization of $d$-wave altermagnetism in \LOMS{} and establish circularly polarized RIXS as a highly symmetry-sensitive spectroscopic framework for detecting magnetic phases that evade conventional probes.}
% ~\cite{Fender2025Altermagnetism,Zhou2025Manipulation}

Symmetry governs the fundamental properties of condensed matter. This is especially true for altermagnetism, a magnetic phase that emerges when two magnetic sublattices of opposite spin orientation are connected by rotational or mirror operations, rather than the translation or inversion symmetries, as it is in the case for conventional antiferromagnets~\cite{Smejkal2022Conventional,Smejkal2022Emerging,Jungwirth2026Symmetry,Cheong2025Altermagnetism,Liu2026Altermagnetism}. Such symmetry considerations induce momentum-dependent, non-relativistic spin polarized electronic bands, and uncompensated chiral magnons~\cite{Chen2025Unconventional,Smejkal2023Chiral,Kravchuk2025Chirala,Beida2025Chiral}. Depending on the underlying crystal structure, these quantities inherit a characteristic $d$, $g$, or $i$ symmetry, giving rise to a host of unconventional macroscopic phenomena --- including piezomagnetic responses, net-magnetization-free spin-current generation, and anomalous Hall and Nernst effects~\cite{Aoyama2024Piezomagnetic,Reichlova2024Observation,Karube2022Observation,GonzalezBetancourt2023Spontaneous,Yang2026AltermagnetDriven,Zhang2025Theory}. Furthermore, the potential for altermagnetism to intertwine with other functional orders, such as unconventional superconductivity or multiferroicity, opens new avenues for multifunctional device architectures.~\cite{Duan2025Antiferroelectric,Gu2025Ferroelectric}.
% ~\cite{Fender2025Altermagnetism,Jin2025Stronga,Cichutek2025Spontaneous,Zhang2025Chirala,Sandratskii2025Direct,Alaei2025Origi,Eto2025Spontaneous,Wu2024ValleyRelated,Gonzalez-Hernandez2021Efficient,Bose2022Tilted,Ma2021Multifunctional,Zhu2024Multipiezo,Wang2026Quantized,Sun2025Proposing,Feng2022anomalous}

Experimentally, the spin-polarized electronic bands of altermagnets can be accessed with \gls*{ARPES} by observing magnetism-induced band splitting or, more definitively, through spin-resolved \gls*{ARPES}~\cite{Krempasky2024Altermagnetic,Ding2024Large,Reimers2024Direct,Jiang2025metallic}. Meanwhile, chiral magnons may be inferred from inelastic neutron scattering, in which magnons of opposite chirality are expected to exhibit distinct dispersions and respond differently to polarized neutrons~\cite{Maier2023Weakcoupling,McClarty2025Observing,Liu2024Chirala,Zhang2025Evidence,Singh2025Chiral}. However, the interpretation of neutron data is sometimes complicated by dipolar interactions, which can similarly generate magnon-band splittings or partially mix opposite-chirality modes~\cite{Sun2025Observationa,Morano2025Absence,Faure2025Altermagnetism,Sears2026Altermagnetic}.

Circularly polarized \gls*{RIXS} has emerged as a potentially more direct probe of chiral magnons, with the chirality encoded in the circular dichroism of the \gls*{RIXS} signals~\cite{Biniskos2025Systematic,Takegami2025Circular,Jost2025Chiral,Channagowdra2025Bialtermagnetism}. Nevertheless, the reliability of this probe remains a subject of intense debate; dichroism can arise from extrinsic effects, such as birefringence~\cite{Nag2025Circular} or the effective time-reversal symmetry breaking inherent to the \gls*{RIXS} process itself~\cite{Biniskos2025Systematic,Furo2025Theory}. Because these mechanisms do not couple directly to the altermagnetic order, they introduce substantial ambiguities, leaving it unclear whether \gls*{RIXS} can serve as a definitive probe of altermagnetism. Here, we resolve this controversy through a systematic study of the altermagnetic candidate \LOMS{}, providing unambiguous experimental evidence of $d$-wave altermagnetism, and elucidating the fundamental coupling mechanism between \gls*{RIXS} and altermagnetism.
% ~\cite{Osumi2024Observation,Fedchenko2024Observation,Liu2024Absencea,Zeng2024Observation}

\section*{Lattice symmetry}

\begin{figure*}
\includegraphics{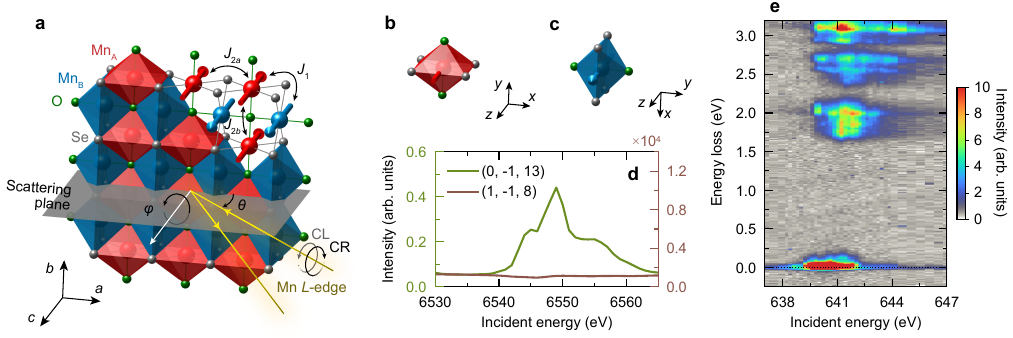}
\caption{\textbf{Crystal symmetry of \LOMS{} and resonant X-ray scattering characterization.} \textbf{a}, Schematic of the \LOMS{} lattice structure, magnetic configuration, and the \acrfull*{RIXS} experimental geometry. Here, $\theta$ and $\phi$ denote the incident and azimuthal angle, respectively; the scattering angle $2\Theta$ was fixed at $154^{\circ}$. Measurements were performed at the Mn $L_3$ edge using both \acrfull*{CL} and \acrfull*{CR} incident polarizations (Methods). \textbf{b},~\textbf{c}, The two symmetry-related sublattices, $A$ and $B$, connected by the $[C_2 \parallel C_{4z}]$ operation. Local Cartesian coordinates for each sublattice are indicated and were utilized in the \acrfull*{ED} calculations. \textbf{d}, Incident energy dependence of Bragg peak intensities measured via the Mn $K$-edge \acrfull*{REXS}. The peak at $\Q{}=(1, -1, 8)$ exhibits minimal energy dependence, characteristic of conventional Bragg peaks, whereas the intensity at $\Q{}=(0, -1, 13)$ is significantly enhanced on resonance due to \acrfull*{ATS} scattering, reflecting the intrinsic symmetry of the material. \textbf{e}, \gls*{RIXS} energy map acquired at 15~K with $\theta=18^{\circ}$ in the \gls*{CL} polarization channel. Features above 1.5~eV energy loss correspond to $dd$ excitations.}
\label{fig:schematic}
\end{figure*}

\LOMS{} is a strongly correlated semiconductor that crystallizes in a quasi-two-dimensional Lieb lattice with the space group $I4/mmm$~\cite{Ni2010Physical,Liu2011Structural,Wei2025LOMS,Asai2026Realization}. The structure comprises distorted MnSe$_4$O$_2$ octahedra forming two distinct Mn sublattices (Mn$_{A}$ and Mn$_{B}$, Fig.~\ref{fig:schematic}a), which are connected by $C_{4z}$ symmetry (Fig.~\ref{fig:schematic}b,~c). Below the transition temperature of $\TN{}\sim160$~K, the system adopts G-type magnetic order with a propagation vector $\k{}=(0, 0, 0)$. The static magnetic moments are collinear along the crystallographic $c$ axis and are strictly antiparallel on the $A$ and $B$ sites, with no evidence of canting or mixing~\cite{Ni2010Physical,Liu2011Structural}. In this ordered state, the Mn sublattices are no longer connected by the $C_{4z}$ operation; instead, they are related by a combined $[C_2 \parallel C_{4z} ]$ symmetry~\cite{Garcia-Gassull2025Microscopic,Chang2025Inverse}. Consequently, both the combined parity-time ($\mathcal{P}\mathcal{T}$) and translation-time ($\tau\mathcal{T}$) symmetries are broken, fulfilling the fundamental criteria for altermagnetism~\cite{Wei2025LOMS}. This is further supported by calculations revealing spin-polarized electronic bands and chiral magnons along the $(H, 0)$ direction, which are absent along the $(H, H)$ direction, reflecting a characteristic $d$-wave symmetry~\cite{Wei2025LOMS,Garcia-Gassull2025Microscopic}. Together, these symmetry analyses predict \LOMS{} as an ideal realization of a $d$-wave altermagnet.

To reconcile this altermagnetic symmetry with the seemingly isotropic character of the material's magnetism, it is necessary to consider the electronic state of the Mn ions. For Mn$^{2+}$ with a $3d^5$ configuration, the orbital degrees of freedom are quenched in the ground state despite the structural distortion of the MnSe$_4$O$_2$ octahedra, effectively recovering $SO(3)$ symmetry for the Mn$^{2+}$ ions. As a result, the Mn$^{2+}$ ions on the two sublattices appear equivalently related by a simple translation operator. Due to this effective translational equivalence, Mn does not contribute to certain structural Bragg reflections allowed by the $I4/mmm$ space group. The $C_{4z}$ symmetry manifests at higher order and becomes observable through resonant X-ray scattering. For example, in \gls*{REXS} at the Mn $K$ edge, the conventional Bragg peak $\Q{}=(1, -1, 8)$, which receives contributions from all ions, shows an intensity that is essentially independent of incident photon energy. In contrast, the intensity of the $\Q{}=(0, -1, 13)$ reflection, which contains only non-Mn scattering factors in the non-resonant case, is dramatically amplified at the resonance (Fig.~\ref{fig:schematic}d). This \gls*{ATS}-like enhancement arises because the resonant process modifies the Mn$^{2+}$ atomic form factor~\cite{Dmitrienko2005Polarization}, allowing the $C_{4z}$ symmetry to emerge and breaking the effective translational connection between the two Mn sublattices. Thus, while the Mn$^{2+}$ ground state appears isotropic, resonant X-ray scattering uncovers the underlying $D_{2h}$ symmetry that drives its altermagnetic behavior.

\section*{Circular dichroism in RIXS}

With the non-translational relationship between the two Mn sublattices established, we performed \gls*{RIXS} measurements at the Mn $L_3$ edge (Methods). The scattering geometry, illustrated in Fig.~\ref{fig:schematic}a, allowed the sample --- featuring an $ab$ cleavage surface --- to be rotated about the structural $c$ axis to vary the azimuthal angle $\phi$. We employed incident photons with \gls*{CL} and \gls*{CR} polarizations without analyzing the scattered photon polarization. Given the two-dimensional nature of the magnetism in \LOMS{}, we fixed the scattering angle $2\Theta=154^{\circ}$ and focused on the in-plane momentum transfer $\Qp{}=(H, K)$ indexed in reciprocal lattice units (r.l.u.). Figure~\ref{fig:schematic}e presents the low-temperature \gls*{RIXS} energy map, which separates cleanly into a low-energy region ($<0.2$~eV), containing the magnetic excitations of interest, and a high-energy sector ($>1.5$~eV), dominated by $dd$ excitations.

\begin{figure*}
\includegraphics{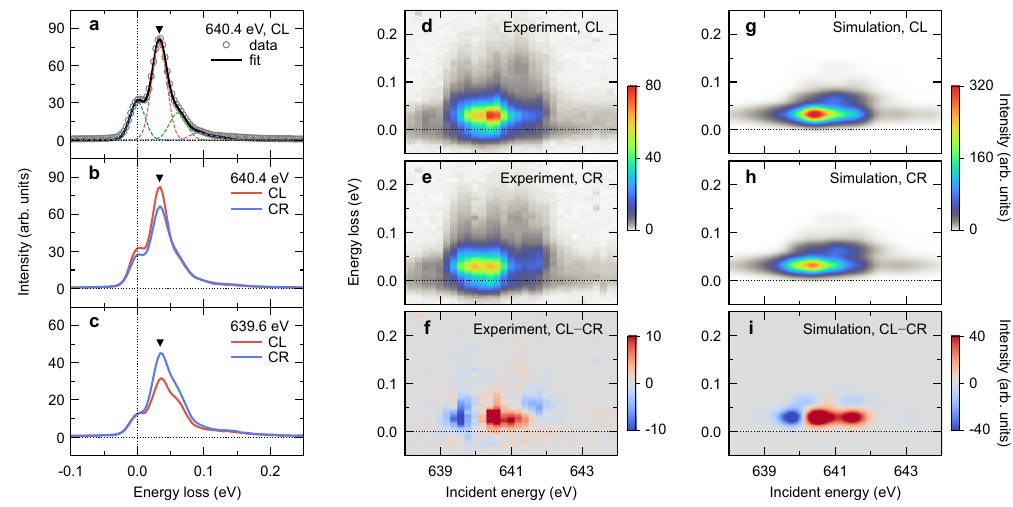}
\caption{\textbf{Circular dichroism in single-magnon excitations.} \textbf{a}, Representative \gls*{RIXS} spectrum in the \gls*{CL} channel. Open circles denote experimental data, and the solid line represents a fit using Gaussian profiles. Individual contributions are indicated by dashed lines: the quasi-elastic line (blue), single-magnon excitation (red), and multi-magnon/phonon contributions (green and purple). \textbf{b},~\textbf{c}, \gls*{RIXS} spectra for the \gls*{CL} and \gls*{CR} channels at the indicated incident energies. Black triangles mark the single-magnon excitations. \textbf{d},~\textbf{e}, Experimental low-energy \gls*{RIXS} energy maps for the two polarization channels. \textbf{f}, Resulting circular dichroism signal, obtained from the intensity difference between the \gls*{CL} and \gls*{CR} channels. \textbf{g}--\textbf{i}, Corresponding \acrfull*{ED} simulations performed under the same experimental geometry. Dotted lines are guides for the eye. All \gls*{RIXS} data were acquired at 15~K with $\theta=18^{\circ}$, corresponding to an in-plane wavevector $\Qp{}\approx (-0.36, 0)$.}
\label{fig:Edep}
\end{figure*}

Figure~\ref{fig:Edep}a displays a representative low-energy \gls*{RIXS} spectrum, which can be decomposed into several components: a quasi-elastic peak centered near zero energy loss, a prominent inelastic peak at $\sim33$~meV, and two higher-energy peaks. Based on polarization analysis, we attribute the 33~meV peak to single-magnon excitations (Supplementary Note~1), while the higher-energy features are likely multi-magnon or phonon excitations --- note that multi-magnon excitations ($\Delta M_s>1$) are fundamentally permitted in \gls*{RIXS}~\cite{Elnaggar2023Magnetic,Li2023Single}. Comparing the \gls*{CL} and \gls*{CR} channels reveals a pronounced circular dichroism in the single-magnon peak (Fig.~\ref{fig:Edep}b). Remarkably, the sign of this dichroism reverses upon tuning the incident photon energy to a lower value (Fig.~\ref{fig:Edep}c). While the intensity of the quasi-elastic line is susceptible to extrinsic factors, such as the sample surface tomography, which can induce artificial dichroism, the dichroism observed in the inelastic signals is intrinsic and robust. Due to the limited statistics and lower intensity of the multi-magnon/phonon signals, the following analysis focuses on the single-magnon contribution, which is directly linked to the underlying altermagnetic order.

Figures~\ref{fig:Edep}d--f map the incident-energy dependence of the low-energy features, which exhibit a resonant enhancement confirming their collective nature. The single-magnon excitations display a circular dichroism that varies with incident photon energy (Fig.~\ref{fig:Edep}f), implying that circularly polarized \gls*{RIXS} does not couple directly to the chiral magnons; rather, the measured cross-section is governed by the complex intermediate states of the resonant process. Given that \LOMS{} is expected to host two altermagnetic domains related by a global reversal of all spin moments (Fig.~\ref{fig:schematic}a--c), we consistently observe two types of domains with opposite dichroism, each extending over several hundred micrometers (Supplementary Note~2). The X-ray beam ($\sim 50~\mu$m horizontal $\times$ $5~\mu$m vertical) is sufficiently small to probe a single domain throughout the measurement, ensuring that the reported dichroism originates from a well-defined altermagnetic domain.

\begin{figure*}
\includegraphics{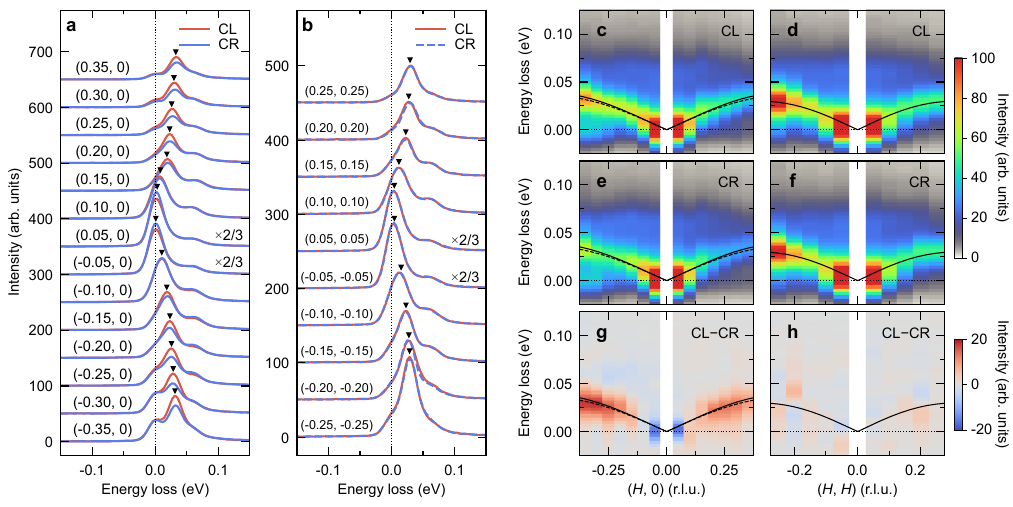}
\caption{\textbf{Magnon dispersion and momentum dependence of circular dichroism.} \textbf{a},~\textbf{b}, \gls*{RIXS} spectra measured at $T=15$~K and $E_{\mathrm{i}}=640.4$~eV in the \gls*{CL} and \gls*{CR} channels along the $(H, 0)$ and $(H, H)$ directions, respectively. Single-magnon excitations are highlighted by black triangles. An intensity offset was applied for clarity. \textbf{c}--\textbf{h}, Momentum-dependent \gls*{RIXS} spectra and the extracted circular dichroism. Solid and dashed curves denote the simulated magnon dispersions of opposite chirality, using exchange parameters $J_1=4.68$~meV, $J_{2a}=1.12$~meV, and $J_{2b}=0.82$~meV. Note the degeneracy of the two branches along the $(H, H)$ direction. Dotted lines are guides for the eye.}
\label{fig:dispersion}
\end{figure*}

To further elucidate the link between circular dichroism and altermagnetism, we examined the momentum dependence of the signal. The single-magnon peak exhibits clear dispersion along both the $(H, 0)$ and $(H, H)$ directions (Fig.~\ref{fig:dispersion}a--f). This dispersion is well reproduced by \gls*{LSWT} calculations using the exchange parameters from a recent inelastic neutron scattering study~\cite{Asai2026Realization} (Fig.~\ref{fig:dispersion}c--f, Methods), further confirming the single-magnon nature of this feature. While simulations predict a small splitting of the opposite-chirality magnon branches ($\sim 2.4$~meV at $\Qp{}=(-0.35, 0)$), which lies beyond the energy resolution of current measurements ($\sim 27$~meV, Supplementary Note~3), a pronounced circular dichroism is observed for magnons propagating along the $(H, 0)$ direction (Fig.~\ref{fig:dispersion}a,~g), whereas it is essentially absent along $(H, H)$ (Fig.~\ref{fig:dispersion}b,~h). This anisotropy constitutes a key signature of $d$-wave altermagnetism. The dichroism diminishes at reduced momentum transfer and reverses sign very close to the zone center (Fig.~\ref{fig:dispersion}g, Supplementary Fig.~9), further highlighting the complex coupling between circularly polarized \gls*{RIXS} and chiral magnons. For completeness, we note that a weak dichroism is also detected in the high-energy $dd$ excitations (Supplementary Fig.~4).

\begin{figure*}
\includegraphics{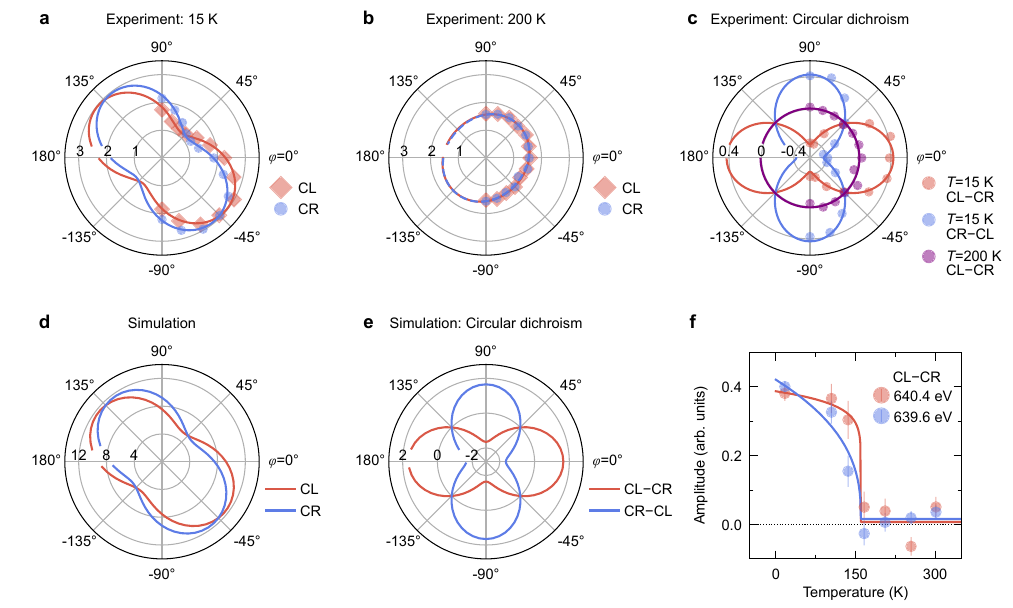}
\caption{\textbf{Azimuthal dependence of circular dichroism.} \textbf{a},~\textbf{b}, Azimuthal angle ($\phi$) dependence of the fitted single-magnon peak amplitudes in the \gls*{CL} and \gls*{CR} channels at the indicated temperatures. Here, $\phi=0^{\circ}$ and $\phi=90^{\circ}$ correspond to the $(-H, 0)$ and $(0, -K)$ directions, respectively. The momentum transfer amplitude $|\Qp{}|$ is fixed at 0.35~r.l.u. \textbf{c}, Corresponding circular dichroism of the single-magnon excitations below (15~K) and above (200~K) \TN{}. Solid lines are guides for the eye. \textbf{d},~\textbf{e}, \gls*{ED} calculations of the azimuthal dependence of the single-magnon peak amplitude and the resulting circular dichroism, respectively. \textbf{f}, Temperature dependence of the circular dichroism at the indicated incident energy for $\Qp{}=(-0.35, 0)$. Solid and dotted lines are guides for the eye. $E_i=640.4$~eV was used for the data and simulations in a--e.}
\label{fig:phi_dep}
\end{figure*}

To determine the symmetry of the dichroism, we measured \gls*{RIXS} spectra as a function of azimuthal angle $\phi$ while keeping the magnitude of the in-plane momentum transfer constant. The magnon intensity for each $\phi$ was extracted via the fitting procedure illustrated in Fig.~\ref{fig:Edep}a, and the results are shown in Fig.~\ref{fig:phi_dep}a,~b. At low temperature, both the \gls*{CL} and \gls*{CR} channels exhibit two-fold symmetry with a small phase shift (Fig.~\ref{fig:phi_dep}a), and their difference yields a dichroism characterized by a well-defined $d$-wave symmetry (Fig.~\ref{fig:phi_dep}c). Upon heating into the paramagnetic state, the azimuthal dependence becomes predominantly isotropic (Fig.~\ref{fig:phi_dep}b), with only marginal two-fold modulations likely arising from residual short-range magnetic correlations. The concurrent disappearance of the dichroism (Fig.~\ref{fig:phi_dep}c,~f) demonstrates that this effect is intimately linked to the low-temperature $d$-wave altermagnetic order. This behavior persists with a different incident photon energy, confirming the robustness of the observation (Supplementary Note~4).

\section*{Microscopic origin of the dichroism}

While our \gls*{RIXS} measurements clearly reveal $d$-wave circular dichroism of magnetic excitations within the ordered phase, the persistence of this signal is initially counterintuitive, as the two chiral magnon branches are nearly degenerate and unresolved. The altermagnetic $[C_2\parallel C_{4z}]$ symmetry pairs the spectral weights of opposite-chirality branches such that the contribution of each magnetic sublattice to one branch is mirrored by the other in the opposite branch.

This apparent contradiction is resolved by the structure of the \gls*{RIXS} operator. The circular dichroism $\Delta I \equiv I^{\text{CL}} - I^{\text{CR}}$ is given by
\begin{equation*}
\Delta I(\mathbf{q}, \omega) \propto
\sum_{\mu,\nu=A,B} \left[\Delta C^{+-}_{\mu\nu} \mathcal{S}_{\mu\nu}^{+-}(\mathbf{q}, \omega)
- \Delta C^{-+}_{\mu\nu} \mathcal{S}_{\mu\nu}^{-+}(\mathbf{q}, \omega)\right],
\end{equation*}
where $\Delta C^{\alpha\beta}_{\mu\nu}$ are polarization-dependent coefficients and $\mathcal{S}_{\mu\nu}^{\alpha\beta}$ are the dynamical structure factors for the two transverse spin-correlation channels. The effective scattering operator can be expanded in local spin operators comprising antisymmetric (rank-1) terms linear in $\mathbf{S}$, symmetric (rank-2) terms quadratic in $\mathbf{S}$, and higher-order contributions \cite{Haverkort2010Theory,Lu2017}. If only rank-1 terms were present, altermagnetic symmetry would enforce a total cancellation of the dichroism when combined with the paired structure factors $\mathcal{S}^{+-}$ and $\mathcal{S}^{-+}$ (Supplementary Notes~7--9). However, rank-2 and higher-order terms lift this cancellation by coupling to single-magnon excitations via the ordered moment with a sublattice-dependent sign, thereby breaking the spectral-weight pairing and generating the observed dichroic signal. The $d$-wave angular dependence is a direct consequence of the altermagnetic symmetry: because the azimuthal rotation axis coincides with the $C_{4z}$ axis, a $90^\circ$ rotation is equivalent to time reversal, strictly reversing the dichroism sign, $\Delta I(\phi) = - \Delta I(\phi + 90^\circ )$. Furthermore, at $\phi=45^\circ$ and $135^\circ$, the scattering plane aligns with the diagonal $(H, H)$ mirror plane, mandating zero dichroism. These constraints enforce the $d$-wave symmetry observed in our \gls*{RIXS} measurements (Fig.~\ref{fig:phi_dep}).

To quantitatively validate these conclusions, we performed \gls*{ED} calculations on two non-interacting atomic sites with local axis related by $C_{4z}$ symmetry (Fig.~\ref{fig:schematic}b,~c). The Hamiltonian accounts for realistic crystal-field splitting, many-body Coulomb interactions, and the intermediate-state core-hole potential, with the \gls*{RIXS} cross section evaluated via the Kramers-Heisenberg equation in the dipole approximation (Methods). While the calculations yield no circular dichroism in the absence of magnetic order (Supplementary Fig.~10), the introduction of antiparallel $z$-axis exchange fields --- consistent with the $[C_2 \parallel C_{4z}]$ symmetry (Fig.~\ref{fig:schematic}b,~c) --- reproduces the experimental results. In this framework, rank-2 and higher-order terms are naturally incorporated, resulting in a robust circular dichroism whose incident-energy (Fig.~\ref{fig:Edep}g--i), azimuthal dependence (Fig.~\ref{fig:phi_dep}d,~e), and momentum dependence (Supplementary Note~6) closely match our observations. Crucially, this dichroism emerges without requiring any splitting between opposite-chirality magnon branches. These results demonstrate that the observed \gls*{RIXS} dichroism is a direct consequence of the lattice and spin symmetries intrinsic to altermagnetism, rather than a signature of chiral-magnon splitting.

In summary, using circularly polarized \gls*{RIXS}, we have uncovered a well-defined circular dichroism in the single-magnon excitations of the strongly correlated Lieb-lattice magnet \LOMS{}. The $d$-wave symmetry and its disappearance in the paramagnetic state provide compelling evidence for its origin in $d$-wave altermagnetism. Together with our \gls*{RIXS} operator analysis and \gls*{ED} calculations, we show that this dichroism arises from the intrinsic symmetries that trigger altermagnetism. Despite the near-degeneracy of the opposite-chirality magnon branches, the nonlinear coupling of the \gls*{RIXS} process to single-magnon excitations generates sublattice-dependent cross-section contributions, leading to uncompensated dichroism. Our findings provide unambiguous evidence for $d$-wave altermagnetism in \LOMS{} and establish circularly polarized \gls*{RIXS} as a powerful, direct probe of altermagnetic symmetry. Notably, this approach is robust against perturbations --- such as dipolar or \gls*{DM} interactions --- that would otherwise introduce or mix magnon chiralities. These results also offer new perspectives on interpreting \gls*{RIXS} circular dichroism for other chiral quasiparticles, such as phonons~\cite{Ueda2023Chiral}.

\section*{Methods}

\subsection*{Sample synthesis}

Polycrystalline \LOMS{} samples were synthesized via a conventional solid-state reaction. High-purity La$_2$O$_3$, Mn, and Se powders were mixed stoichiometrically and pressed into pellets. The pellets were sealed in evacuated quartz tubes and sintered at 950$^{\circ}$C for 24~hours. Single crystals were grown by loading the polycrystalline \LOMS{} precursor into an alumina crucible, which was then sealed within an argon-filled iron crucible. The assembly was sintered at 1430$^{\circ}$C for 3~hours and slowly cooled to 1350$^{\circ}$C at a rate of 1.5$^{\circ}$C/hr. Black single crystals were harvested from the bottom of the alumina crucible. The phase purity and crystallinity of the samples were verified using laboratory-based X-ray diffraction (Supplementary Fig.~12).

\subsection*{Resonant X-ray scattering experiments}

\gls*{RIXS} and \gls*{REXS} experiments were performed at the I21 and I16 beamlines of the Diamond Light Source (UK), respectively~\cite{Zhou2022I21,Collins2010Diamond}. A single crystal with the $c$ axis as the surface normal was utilized for both experiments. For the I21 \gls*{RIXS} experiment, the incident photon energy was tuned to the Mn $L_3$ edge using both \gls*{CL} and \gls*{CR} polarizations; the polarization of the scattered photons was not analyzed. The combined energy resolution was approximately 27~meV. All \gls*{RIXS} data presented were normalized to the incident flux, monitored via the focusing mirror current, to ensure consistency between different polarizations. While self-absorption effects were not explicitly corrected, the sample's $ab$ cleavage surface ensures that both self-absorption and the photon footprint remain invariant under rotations of the azimuthal angle $\phi$, thus maintaining constant scattering efficiency throughout the azimuthal dependence. During azimuthal and temperature-dependent measurements, sample positioning was monitored via an optical camera to ensure that the same spot was probed. Given the large characteristic size of the altermagnetic domains, we conclude that the measurements remained within a single domain throughout the process.

\subsection*{Linear spin wave analysis}

We consider a Heisenberg model for the Mn layer with two sites per unit cell, $A$ at $\mathbf r_A=(\tfrac12,0)$ (Fig.~\ref{fig:schematic}b) and $B$ at $\mathbf r_B=(0,\tfrac12)$ (Fig.~\ref{fig:schematic}c). The Hamiltonian reads
\begin{equation*}
H = J_1 \sum_{\langle i,j\rangle} \mathbf S_i \cdot \mathbf S_j +
    \sum_{\langle\!\langle i,j\rangle\!\rangle_{\mu}^{\alpha}} J_2^{\alpha,\mu} \mathbf S_i \cdot \mathbf S_j.
\end{equation*}
Here, $J_1$ is the nearest-neighbor exchange coupling between sites on different sublattices, and $J_2^{\alpha,\mu}$ are the anisotropic second-neighbor exchanges within the same sublattice $\mu \in \{A, B\}$ along direction $\alpha \in \{x,y\}$. $J_2^{x,A}=J_2^{y,B}=J_{2a}$ and $J_2^{y,A}=J_2^{x,B}=J_{2b}$. We use $J_1=4.68, J_{2a}=1.12, J_{2b}=0.82$ in units of meV~\cite{Asai2026Realization}. Performing the standard \gls*{LSWT} analysis, we obtain the single magnon dispersion
\begin{equation*}
\omega_{\pm}(\mathbf k)=
\sqrt{\left(\frac{\epsilon_A(\mathbf k)+\epsilon_B(\mathbf k)}{2}\right)^2-\Delta(\mathbf k)^2} \pm
\frac{\epsilon_A(\mathbf k)-\epsilon_B(\mathbf k)}{2},
\end{equation*}
with
\begin{equation*}
\begin{split}
\epsilon_A(\mathbf{k}) &= S \left[ 4J_1 + 2J_{2a}(\cos k_x - 1) + 2J_{2b}(\cos k_y - 1) %+ 4J_3(\cos k_x \cos k_y - 1)
\right], \\
\epsilon_B(\mathbf{k}) &= S \left[ 4J_1 + 2J_{2b}(\cos k_x - 1) + 2J_{2a}(\cos k_y - 1) %+ 4J_3(\cos k_x \cos k_y - 1)
\right], \\
\Delta(\mathbf{k}) &= 4 S J_1 \cos \frac{k_x}{2} \cos \frac{k_y}{2}.
\end{split}
\end{equation*}
The corresponding transverse dynamical structure factors, Bogoliubov coefficients, and symmetry relations that pair the two opposite-chirality branches are given in Supplementary Note~7.

\subsection*{ED calculations}

\gls*{ED} calculations were performed using the EDRIXS software~\cite{Wang2019EDRIXS}. The model explicitly accounts for onsite electron-electron interactions and the core-hole potential, which govern the intermediate state of the scattering process. We applied a \gls*{CNN} approach to determine the electronic parameters that best describe our experimental results (see Supplementary Note~5 for details). The optimized crystal-field parameters and the Slater integrals for the Coulomb interactions and core-hole potentials were found to be (in units of eV, to two decimals) $E_{x^2-y^2}(A_g)=0.76$, $E_{3z^2-r^2}(A_g)=-0.42$, $E_{xz}(B_{2g})=0.25$, $E_{yz}(B_{3g})=-0.34$, $E_{xy}(B_{1g})=-0.25$, $F^0_{dd}=0.33$, $F^2_{dd}=7.47$, $F^4_{dd}=2.84$, $F^0_{dp}=0.23$, $F^2_{dp}=3.93$, $G^1_{dp}=2.54$, and $G^3_{dp}=1.45$. A mixing term of $-0.02$~eV was included between the $x^2-y^2$ and $3z^2-r^2$ orbitals, as both belong to the $A_g$ irreducible representation. The spin-orbit coupling constants for the initial and intermediate states were fixed at the atomic value of 0.04 and 0.053~eV, respectively. As described in the main text, the calculations were performed on two non-interacting atomic sites representing the two Mn sublattices. Both sites share identical crystal-field parameters and Slater integrals, reflecting the equivalence of their local MnSe$_4$O$_2$ octahedra; however, opposite exchange fields were applied to model the G-type magnetic order. To evaluate the \gls*{RIXS} cross-sections, the photon polarization vectors were transformed from the global coordinate system to the local coordinates of each site, which are related by $C_{4z}$ symmetry (Fig.~\ref{fig:schematic}a--c). The experimental geometry, defined by the angles $2\Theta$, $\theta$, and $\phi$, was explicitly incorporated into the definition of these polarization vectors.

\section*{Data availability}

All data that support the findings of this study have been deposited in the Zenodo database with the access code [to be assigned].

\section*{References}
\bibliographystyle{naturemag}
\bibliography{refs}
\clearpage

\section*{Acknowledgements}

We thank Sergey V. Streltsov and Changle Liu for valuable discussions. This work was supported by the National Key R\&D Program of China (Grant No.~2024YFA1408301 (Y.S.), 2022YFA1403000 (Y.L.), 2021YFA1400300 (Y.W.L.)) and the National Natural Science Foundation of China (Grant No.~12574139 (Y.S.), 12274207 (Y.L.), 12425403 (Y.W.L.)). Q.W. acknowledges support from the Research Grants Council of Hong Kong (Grant No.~CUHK 24306223), the Guangdong Provincial Quantum Science Strategic Initiative (Grant No.~GDZX2401012), and the 1+1+1 CUHK-CUHK(SZ)-GDSTC Joint Collaboration Fund (Grant No.~2025A0505000079). V.C.M. acknowledges support from the Swiss National Science Foundation (Grant No.~200021\_219950). The REXS and RIXS experiments were performed at beamlines I16 and I21 at the Diamond Light Source, respectively (proposals MM39123 and MM40708).

\section*{Author contributions}

Y.S. conceived the project. G.Z., X.Y., S.T., M.P., and M.Y. synthesized and characterized the samples. G.Z. and Y.S. performed the X-ray measurements with support from J.C., R.S., M.G.-F., and A.B. G.Z., X.Y., V.C.M., Z.P., X.Q., D.G.M., Q.W., Y.L., Y.S., and Y.W.L. interpreted the data. Y.H.L., S.T.L., Q.W., Y.L., and Y.S. performed the calculations. The paper was written by G.Z., Q.W., Y.L., and Y.S. with input from all co-authors.

Correspondence and requests for materials should be addressed to Qisi Wang, Yi Lu, or Yao Shen.

\section*{Competing interests}

The authors declare no competing interests.

\clearpage

\end{document}